# The Padul normal fault activity constrained by GPS data: brittle extension orthogonal to folding in the central Betic Cordillera

**Antonio J. Gil** [a,b], **Jesús Galindo-Zaldívar** [c,d, *], **Carlos Sanz de Galdeano** [d], **Mª Jesús Borque** [a,b], **Alberto Sánchez-Alzola** [e], **Manuel Martinez-Martos** [d], **Pedro Alfaro** [f]

[a] *Dpto. Ingeniería Cartográfica, Geodésica y Fotogrametría, Universidad de Jaén, 23071 Jaén, Spain.* ajgil@ujaen.es; mborque@ujaen.es;

[b] *Centro de Estudios Avanzados en Ciencias de la Tierra (CEACTierra), Universidad de Jaén. 23071 Jaén, Spain.*

[c] *Dpto. Geodinámica, Universidad de Granada, 18071 Granada, Spain.* jgalindo@ugr.es

[d] *IACT, CSIC-Universidad de Granada, 18071 Granada, Spain.* csanz@ugr.es; manuelmm@ugr.es

[e] *Dpto. Estadística e Investigación Operativa, Universidad de Cádiz. 11405 Cádiz, Spain.* alberto.sanchez@uca.es

[f] *Dpto. Ciencias de la Tierra y Medio Ambiente, Universidad de Alicante, Alicante, Spain.* pedro.alfaro@ua.es

*Corresponding author: TF: +34958243349 FAX: +34958248527 Email: jgalindo@ugr.es

## Abstract

The Padul Fault is located in the Central Betic Cordillera, formed in the framework of the NW-SE Eurasian-African plate convergence. In the Internal Zone, large E-W to NE-SW folds of western Sierra Nevada accommodated the greatest NW-SE shortening and uplift of the cordillera. However, GPS networks reveal a present-day dominant E-W to NE-SW extensional setting at surface. The Padul Fault is the most relevant and best exposed active normal fault that accommodates most of the NE-SW extension of the Central

Betics. This WSW-wards dipping fault, formed by several segments of up to 7 km maximum length, favored the uplift of the Sierra Nevada footwall away from the Padul graben hanging wall. A non-permanent GPS network installed in 1999 constrains an average horizontal extensional rate of 0.5 mm/yr in N66ºE direction. The fault length suggests that a (maximum) 6 magnitude earthquake may be expected, but the absence of instrumental or historical seismic events would indicate that fault activity occurs at least partially by creep. Striae on fault surfaces evidence normal-sinistral kinematics, suggesting that the Padul Fault may have been a main transfer fault of the westernmost end of the Sierra Nevada antiform. Nevertheless, GPS results evidence: (1) shortening in the Sierra Nevada antiform is in its latest stages, and (2) the present-day fault shows normal with minor oblique dextral displacements. The recent change in Padul fault kinematics will be related to the present-day dominance of the ENE-WSW regional extension versus ~NNW-SSE shortening that produced the uplift and northwestwards displacement of Sierra Nevada antiform. This region illustrates the importance of heterogeneous brittle extensional tectonics in the latest uplift stages of compressional orogens, as well as the interaction of folding during the development of faults at shallow crustal levels.

# 1. Introduction

The development of normal faults at shallow crustal levels accompanies shortening during the latest stages of cordillera uplift (Hodges et al., 1998; Galindo-Zaldívar et al., 2003; McDermott et al., 2015). The Eurasian-African plate convergence has a NW-SE trend in the westernmost Mediterranean (Nocquet, 2012), where the Betic-Rif Cordillera is located. In the Central Betic Cordillera, present-day deformation is split into E-W to NE-SW folding accommodating ~NNW-SSE shortening, and normal faulting accommodating ~ENE-WSW extension (Galindo-Zaldívar, 2003, 2015; Ruano et al., 2004).

The most important fault sets in the Betic Cordillera (Fig. 1), developed from the late Miocene onward, show NNE-SSW (mainly affecting the eastern part of the cordillera) and NW-SE orientations (Sanz de Galdeano, 1983). These faults interact with large folds promoting the formation of the intramontane Neogene basins. Normal faults contributed together with the development of large antiforms to the uplift of the Sierra Nevada and the relative subsidence of the Granada Basin (Figs. 1 and 2). The total throw, considering multiple faults, is of the order of 5000 m. These faults formed under NNW-SSE compression (from NW-SE to NNE-SSW) combined with orthogonal extension, displacing upper Neogene to Quaternary sediments. Therefore, several of the remain active producing seismicity (Sanz de Galdeano et al., 2003; Sanz de Galdeano et al., 2012). Sierra Nevada constitutes a large antiformal structure (Fig. 1) formed since the late Serravallian, with a roughly E-W orientation that westwards becomes NE-SW at the southwestern antiformal end (Pedrera et al., 2012).

The development of folds, the elastic behavior of faults, and the rheological heterogeneities in deformation zones can disturb local stresses, as revealed by the earthquake focal mechanisms of the Central Betics and paleostress analysis (Galindo-Zaldívar et al., 1999). Local GPS networks improve the determination of present-day fault kinematics and rates obtained from geological observations, contributing to better understand the variability and behavior of active faults. GPS processing methods such as Precise Point Positioning (PPP) (Zumberge et al., 1997) stand as valuable tools for investigating geodynamical processes at the millimetric level (Larson et al., 2004; Smith et al., 2004; Kouba 2005; Hreinsdóttir et al., 2006).

The aim of this research is to analyze the present-day and the recent behavior of the Padul Fault from geological and GPS data (period 1999-2012) in order to determine its relevance in the framework of regional deformation and the development of the Sierra Nevada antiform. This research provides new data on the present-day tectonic activity of the Padul Fault, the best exposed fault of the Central Betics, and contributes to discussion of its seismic behavior. Its activity is analyzed in the context of fault and fold interaction accommodating roughly simultaneous extensional and compressional deformation.

## 2. Geological setting

The Betic Cordillera, together with the Rif (Fig. 1), belongs to the Alpine belt of the Western Mediterranean, forming the Gibraltar Arc in the Eurasian and African plate boundary. The Alboran Sea is the main Neogene basin located in the central part of the orogen. The Internal Zone is formed by several superposed metamorphic complexes, the main ones (from bottom to top) being the Nevado-Filábride, the Alpujárride and the Malaguide. Flysch Units crop out discontinuously along the contact between the Internal and the External Zones. Sierra Nevada (Fig. 1) belongs to the Internal Zone and is formed by Nevado-Filabride and Alpujarride complexes, composed of graphite-bearing schist, quartz- feldspar rich schist and marbles of Paleozoic to Triassic ages that have undergone HP/LT alpine metamorphism (Monié et al., 1991). These complexes are separated by the low-angle normal Mecina fault, active during the Early to Middle Miocene (Galindo-Zaldívar et al., 1989; Jabaloy et al., 1992). The Late Miocene to present-day NW-SE to N-S shortening in the Betic Cordillera has been accommodated by large E-W to NE-SW folds and several strike-slip and normal fault sets (Sanz de Galdeano, 1983; Galindo-Zaldívar et al., 2003) producing the main reliefs and also the development of intramontane sedimentary basins (Braga et al., 2003; Sanz de Galdeano and Alfaro, 2004). Sierra Nevada constitute the largest and highest antiformal structure of the Betic Cordillera. The main fault sets have NW-SE, NE-SW and E-W orientations and were active since the Miocene (Sanz de Galdeano, 1993). N-S compression and orthogonal extension paleostress fields was related to NW-SE dextral and NE-SW sinistral faults. Anticlockwise rotation of compression to NW-SE produced the development of E-W dextral faults and the reactivation of NW-SE faults that became normal (Sanz de Galdeano, 1993).

Seismicity of the region is irregularly distributed and belongs to the broad active deformation band, more than 300 km wide, that separates Eurasian and Africa in the westernmost Mediterranean (Udías and Buforn, 1991; Galindo-Zaldivar et al., 1993). The seismic activity in the 1990-2016 period obtained from I.G.N. database (www.ign.es) reveals a high concentration of epicenters in the Neogene-Quaternary Granada Basin, located west of Sierra Nevada (Fig. 1), mainly related to NW-SE active normal faults (Sanz de Galedano et al., 2003; 2012). However, seismicity decreases southeastward, in Sierra Nevada, Sierra de Lujar and Sierra de la Almijara and also in the Padul graben, with scarce widespread activity (Fig. 1).

A regional GPS network was deployed by the Topo-Iberia project (Garate et al., 2015). The available data support the present-day W and WSW displacements of the Internal Zones of the Central Betic Cordillera with respect to the relatively stable Iberian foreland (Galindo-Zaldívar et al., 2015). This displacement field evidences an E-W to ENE-WSW extensional deformation in the westernmost part of Sierra Nevada, in addition to a locally very moderate NW-SE shortening along the westernmost part of the Sierra Nevada antiform.

### 3. The Padul Fault

The Padul Fault (Figs. 1, 2, 3 and 4), located in the Internal Zone of the Betic Cordillera, is the most remarkable normal fault of this region from a geomorphic point of view. This active fault separates the highest reliefs of Sierra Nevada (footwall) from the Granada basin that in this sector is represented by the Padul graben (Sanz de Galdeano and Alfaro, 2004), developing a spectacular mountain front (Figs. 3A and 3B) sculpted in Alpujarride marbles (Lhénaff, 1965; Calvache et al., 1997). This NW-SE fault is formed by a northern segment of 5.25 km and a southern segment of 7 km connected by a 1.5 km relay fault (Figs. 1 and 2). The throw of the fault is substantial —over 800 m in its central part— (Santanach et al., 1980).

The fault surfaces have a variable dip to the SW, ranging from 60-65º to less than 20º. Its displacements are predominantly normal, although in many places there is a slight to moderate left lateral component (Fig. 3C). In the hanging wall block, the Padul graben formed since the end of the Miocene, and it is filled by Late Tortonian calcarenites and Messinian marls. During the Pliocene and the Quaternary, the basin was filled by mainly alluvial fans and peat deposits, which merge laterally with lacustrine deposits towards the center of the basin (Domingo-García et al., 1983; Delgado et al., 2002). The most recent alluvial deposits, dating from Late Pliocene to Pleistocene in age, can be grouped into various units (Fig. 2). At the base is a Plio-Pleistocene unit called the Red Formation, made up of reddish-brown clays, sands and alluvial conglomerates. Overlying these rocks are conglomerates with grey and reddish boulders deposited by the Torrente river (Nigüelas Formation). At the top of the sequence, the alluvial fans of the Late Pleistocene-Holocene are found, developed along the north-eastern edge of the basin.

The fault has a continuous scarp developed in the contact with the footwall basement, and well preserved small scarps (1-2 m high) in recent sediments, in some cases showing a complex geometry locally including dextral striae (Fig. 3D). In the sector where the GPS-network is installed, the complex geometry of the Padul fault zone has produced a wide variety of dips in the Plio-Quaternary alluvial deposits (Fig. 4). The older Plio-Pleistocene alluvial sediments from the Red Formation lie directly over low-angle fault planes, showing a back-tilting of 5°-45° towards the NE (Fig. 4). These alluvial sediments cover the triangular facets of the mountain front (Fig. 4). Contrariwise, the Late Pleistocene and Holocene alluvial unconsolidated sediments dip basinwards. Even though the Padul Fault shows evidence of paleoseismic activity (Alfaro et al., 2001), there is no significant related seismicity in the historical and instrumental record.

Southwest of Sierra Nevada, a set of roughly E-W oriented normal faults offers evidence of recent activity (Fig. 1, 2 and 5): fresh fault scarps and/or recent sediments are deposited on the hanging wall accommodating a roughly N-S regional extension.

## 4. GPS observations and data processing

In 1999, an eight-site network was established over the Padul Fault (Fig. 2B) (Ruiz et al., 2003). The location of each individual site was selected to provide good coverage of the main geological structure in both fault blocks. They were installed on exposed rocks using a self-centering mounting device that guarantees the reinstallation of the antenna exactly at the same horizontal position in each campaign. Thus far five GPS field campaigns were carried out —in March 1999, May 2000, July and September 2011, and finally November 2012— reoccupying the sites for a minimum period of five hours during three days in the older campaigns (1999 and 2000) and 72 hours of continuous observation in the more recent ones (2011 and 2012).

In our work the 6.2 version of GIPSY-OASIS software (Gregorius, 1996) was used together with the Precise Point Positioning (PPP) method and the zero-ambiguity resolution strategy described in Bertiger et al. (2010). An identical standard procedure was applied for all campaigns as shown in Sánchez-Alzola et al. (2014): JPL final ephemeris and Pole products were used from a JPL server in a homogenous IGS08 reference frame. The FES2004 ocean tide loading model (Lyard et al., 2006) was also applied. Additionally, the hydrostatic and wet components of the zenith tropospheric delay were included, a 10° cut-off angle was set up, and a calibration file was used to correct the Antenna Phase Center.

## 5. Padul Fault active displacements from GPS data

### 5.1. GPS station position time series

The best trend line was fitted to the time series of positions for each station using linear regression. Figure 6 shows the GPS position time series for Padul Fault sites in the IGS08 frame. The 3PAD time series has an additional observation campaign in July 2001 because this site belongs to the Granada Basin monitoring network observed in that year (Gil et al., 2002). Older campaigns have more dispersion due to the poorer quality of the support files; accuracy is better in the campaigns where the JPL products feature modern data analysis.

Table 1 shows the velocity field in IGS08 reference frame in the horizontal components with respect to the best-fit trend line, instead of the associated formal uncertainties derived from the processing software. Associated errors are computed considering the standard deviation of the slope component. These velocities are consistent with the theoretical tectonic movement in the area with deviations between 3.0-9.0 mm/yr, similar to those of other studies with GPS episodic campaigns (Perez-Peña et al., 2010a; Rayan et al., 2010). Given the small number of solutions, we did not eliminate any observation unless we had objective reasons such as an incorrect installation or unreliable solution.

## 5.2. GPS-derived velocity field

Figure 7 shows the velocity field in IGS08 reference frame of the Padul Fault network, using solutions from the five campaigns between 1999 and 2012. The large time interval, spanning about 13 years, seems suitable to reduce slope error in episodic campaigns and allowed us to compute reliable estimations of velocity for our sites. Position time series longer than 7.5-8 years long allow for reliable deformation rates in horizontal motion with episodic GPS campaigns (Sanli et al., 2012; Gianniou and Stavropoulou, 2016). The standard error ellipses are based on the formal error derived from the linear trend estimation and deviations. We did not consider the existence of any change in site position during the campaign (the effects of secular motions are negligible). The dispersion is attributed to white noise due to the errors in acquired observations. Figure 8 shows the residual velocity field computed with respect to the constant average velocity of 4PAD, 5PAD, 7PAD and 8PAD sites. This model is based on the theoretical fault line between these points situated in the footwall, and 1PAD, 2PAD, 3PAD and 6PAD sites located in the hanging wall. Angles and magnitudes of the vectors in the downthrown hanging wall sites are consistent with the fault activity and shows a southwestwards relative displacement with respect to the footwall block. A rate of roughly 0.5 mm/yr of extension in N66ºE was determined for the region.

## 6. Discussion

The present-day activity of the Padul Fault in the two sectors best constrained by the GPS network (Figs. 2, 7 and 8; Table 1) has an approximate horizontal rate of 0.5 mm/yr (0.44 to 0.55 mm/yr; 0.4 to 0.5 mm/yr of E-W component and 0.2 mm/yr of N-S component) horizontal extension in a N66ºE trend. These high deformation rates in very small distances, sometimes less than 500 m (e.g. 2PAD and 5PAD sites), in addition to the field geological observations, supports that most of the deformation occurs near the relatively narrow damage zone associated to the fault. Considering that the average fault dip is 55º, the average displacement on the fault surface should be of 0.78 mm/yr. These geodetic results, obtained for the first time in the Padul Fault, are slightly higher than those (0.16 to 0.35 mm/yr) deduced from stratigraphic markers (Sanz de Galdeano et al., 2012).

Taking into account that the largest fault segment is about 7 km in length, a hypothetical earthquake of M=6 might be produced by this segment with a total slip of 0.12 m, according to the Wells and Coppersmith (1994) relationships. With the measured GPS rates, this slip is accumulated every 154 years. However, the historical and instrumental seismic records (www.ign.es) do not provide any evidence of large earthquakes in this sector, at least in the last 500 yrs, suggesting that the fault is now unlocked. Despite geological evidence of seismic events in the area (Alfaro et al., 2001), these results suggest that this major fault may have a mixed behavior, including seismic and aseismic deformation.

The activity of the Padul Fault evidenced by the local GPS network (Figs. 2, 7 and 8) may be considered in the framework of the regional deformation (Fig. 1) established by the Topo-Iberia network (Galindo-Zaldívar et al., 2015; Gárate et al., 2015). The relative displacement of the PALM site, located far in the hanging wall southwestward of the fault, with respect to the NEVA site, located far in the Sierra Nevada footwall, indicated a 1.43 mm/yr horizontal extension in N22ºE orientation with a 0.53 mm/yr eastward component and 1.33 mm/yr southward component. These data suggest that the Padul Fault, with a rate of 0.5 mm/yr, accommodated most of the ENE-WSW regional extension southwestward of Sierra Nevada.

According to the regional GPS network (NEVA and PALM GPS sites), the deduced N-S regional extension (1.13 mm/yr) is only partially justified by the Padul Fault, and would entail the activity of other brittle structures. This N-S extension is related to a set of roughly E-W normal faults, of smaller sizes than the Padul Fault and with widespread deformation in the entire region, such as those observed in several localities southwest of Sierra Nevada (Figs. 1, 2 and 5).

Simultaneous compression and orthogonal extension generally determine the presence of strike-slip faulting (Anderson, 1951). However, while compression produces folding, extension should be accommodated in the same regional deformation field by the development of normal faults. This setting is favored in regions with rheological layered crusts, where large crustal detachment levels may develop, as in the Betic Cordillera (Galindo-Zaldívar et al., 1997, 2003). Although the interaction of folds and faults has been accurately studied in fold-and-thrust belts (Chapple, 1978; Davis et al., 1983), there are no well-constrained examples of major normal faults driven by the interaction of nearby fold development. This process occurs in the complex tectonic scenario of the Granada Basin area, one of the regions of the Betic Cordillera with most intense seismic activity (Morales et al., 1997; Sanz de Galdeano et al., 2003). The NW-SE Eurasian-African plate tectonic convergence (Nocquet, 2012) is resolved in the central part of the Internal Zones of the Betic Cordillera by a regional NNW-SSE shortening that favors the uplift of the Cordillera during the development of the Sierra Nevada antiform, and orthogonal NE-SW extension accommodated by the WSW-dipping normal Padul fault zone (Fig. 9).

The comparison of geologic and geodetic data reveals a slightly different behavior of the Padul Fault during the Plio-Quaternary and at Present (Fig. 9). The striae in the well-exposed fault surfaces separating Alpujarride marbles in the footwall and Plio-Quaternary alluvial fans in the hanging wall reveal a normal sinistral component of movement (Figs. 2 and 3C). These kinematics may be a consequence of the activity of the Padul Fault as a transfer fault that accommodated higher shortening rates in the northeastern block (deformed by the large antiform of Sierra Nevada) than in the southwestern block (Fig. 9), producing a local deviated stress field.

Nevertheless, GPS data (Fig. 8) determine a present purely normal, or even slightly normal-dextral slip. Although Perez-Peña et al. (2010b) evoke an active elevation of Sierra Nevada, the present-day decrease in uplift rates supported by regional GPS data (Galindo-Zaldívar et al., 2015) may influence the change in kinematics of the Padul fault —formerly transtensional sinistral, but now mainly pure-normal or slightly normal dextral (Fig. 9), which is in agreement with present regional stresses (e.g., Pedrera et al., 2011). The recent change in kinematics due to the interaction of folds and faults evidences the variability of the local stress fields both in space and time, and the occurrence of poorly congruent earthquake focal mechanisms (Galindo-Zaldívar et al., 1999).

## 7. Conclusions

The Padul Fault, located southwest of Sierra Nevada is one of the main active faults of the south Iberian Peninsula and accommodates most of the ENE-WSW extension in this sector of the Central Betic Cordillera. This NW-SE oriented southwestwards dipping normal fault separates Alpujarride marbles of the footwall from the Padul Graben Plio-Quaternary alluvial fan sediments in the hanging wall. GPS data determines an extensional horizontal rate of 0.5 mm/yr in a N66ºE trend, slightly higher than those deduced previously from stratigraphic markers. Padul Fault probably has a significant part of aseismic deformation, without remarkable seismic activity in the last 500 yr, although more data covering a broad time period are necessary in order to quantify the percentages of aseismic and seismic deformation of this active fault.

Strain partitioning occurs in the central Betics; while compression is accommodated by folding, extension produced the simultaneous activity of NW-SE and E-W extensional faults (Fig. 9). The decrease activity of the Sierra Nevada antiform probably influenced changes in fault kinematics from normal sinistral to purely normal or slightly normal dextral. In this setting, the Padul Fault can be considered as a transfer fault accommodating the deformation at the southwestern end of the Sierra Nevada antiform. The Padul Fault and Sierra Nevada antiform constitute an excellent example of the interaction between faults and folds that determines relief uplift in the latest stages of shortening of the Betic Cordillera.

## Acknowledgements

We acknowledge the comments of Dr. A. Pedrera, two anonymous reviewers and the editor Dr. Kelin Wang which have improved the quality of this paper. This research was funded by PAIUJA 2017/2018 UJA2016/00086/001 project, CGL2016-80687-R AEI/FEDER, UE project and RNM148 and RNM282 research groups of Junta de Andalucía. We also thank J. Sanz de Galdeano for A and B photographs of Figure 3. Some figures were generated using GMT (Wessel and Smith, 2013).

**Figure captions**

Fig.1. Regional geological setting of the Padul Fault. A. Betic Cordillera and Rif placed in between Iberian and African forelands. B. Geological map of the central part of the Betic Cordillera. Epicenters (depth <30 km; 1990 to 2016) from IGN database (www.ign.es).

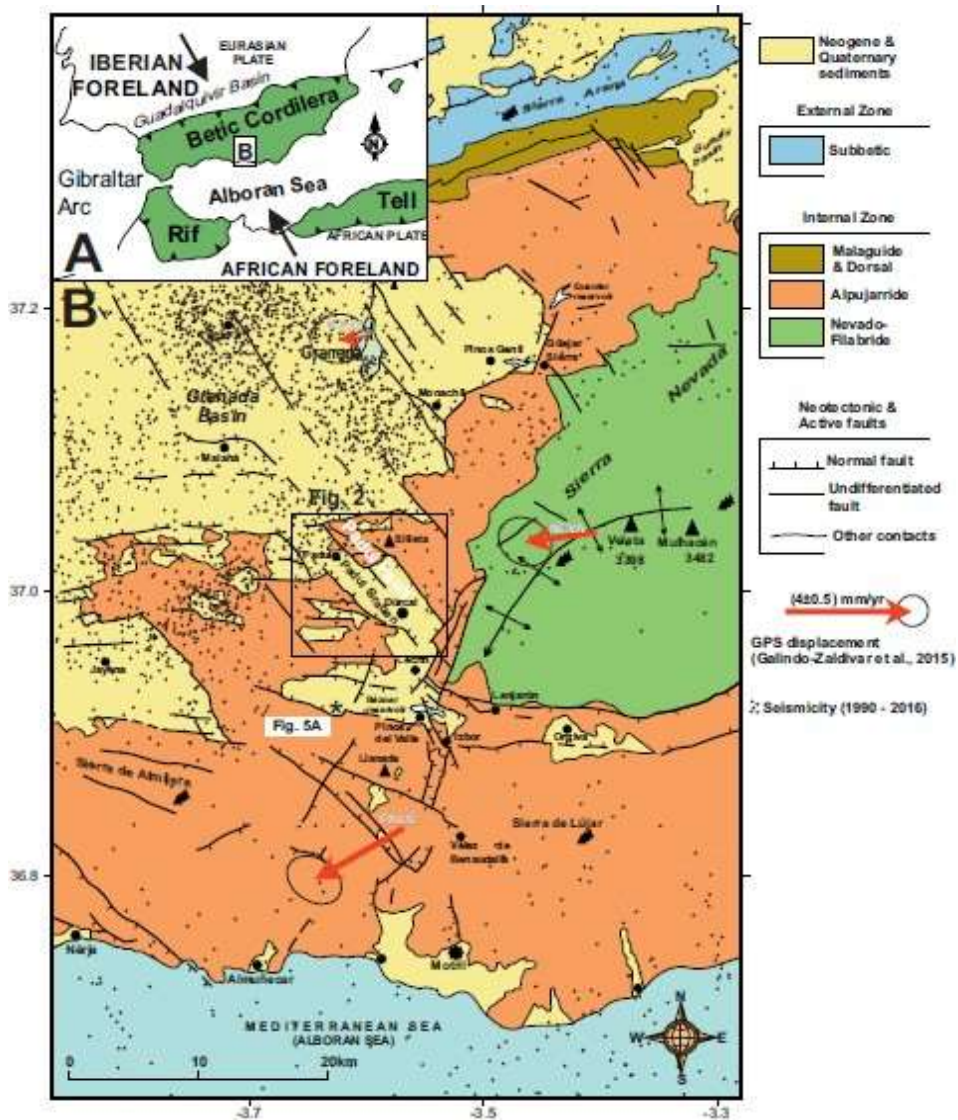

Fig. 2. The Padul Fault. Detailed geological maps of the fault (A), including position of GPS network and straie in its central part (B). Location is marked in Fig. 1.

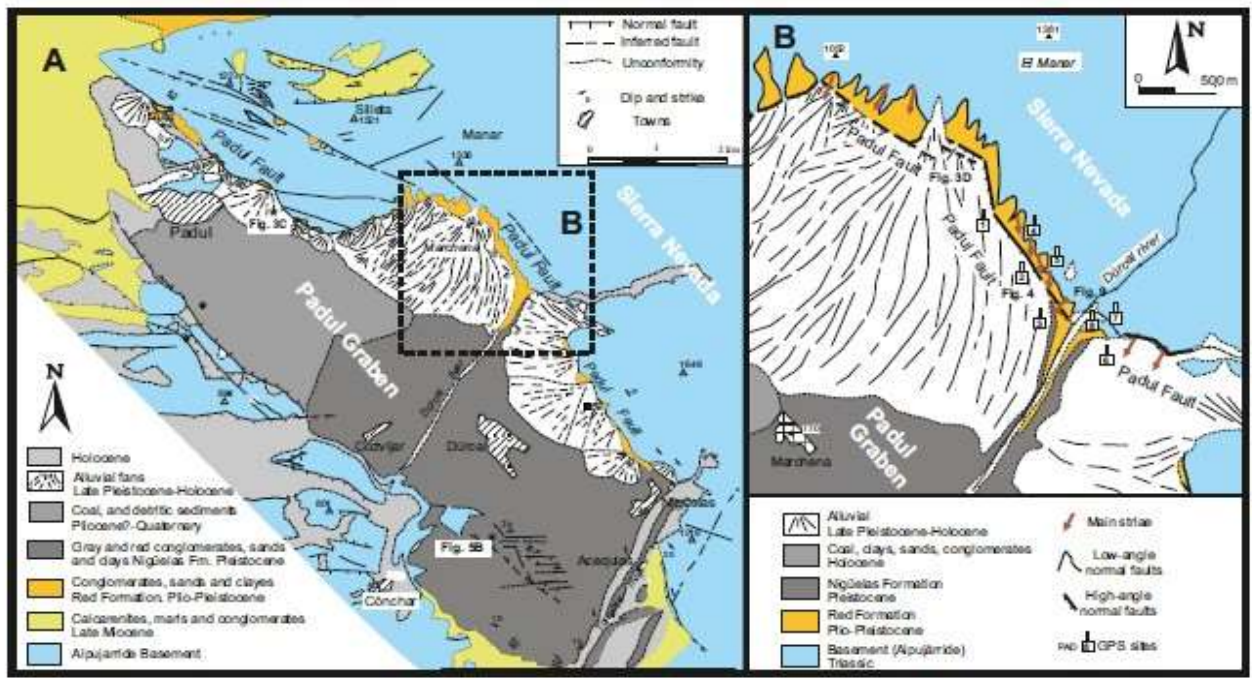

Fig. 3. Field structures of the Padul Fault. Aerial view (A and B). Main fault surface separating Alpujarride Triassic marbles and Plio-Quaternary alluvial fan deposits with normal sinistral striae (C) in a segment indicated in Fig. 2A. Recent dextral striae in alluvial fan deposits (D) in a segment indicated in Fig. 2B.

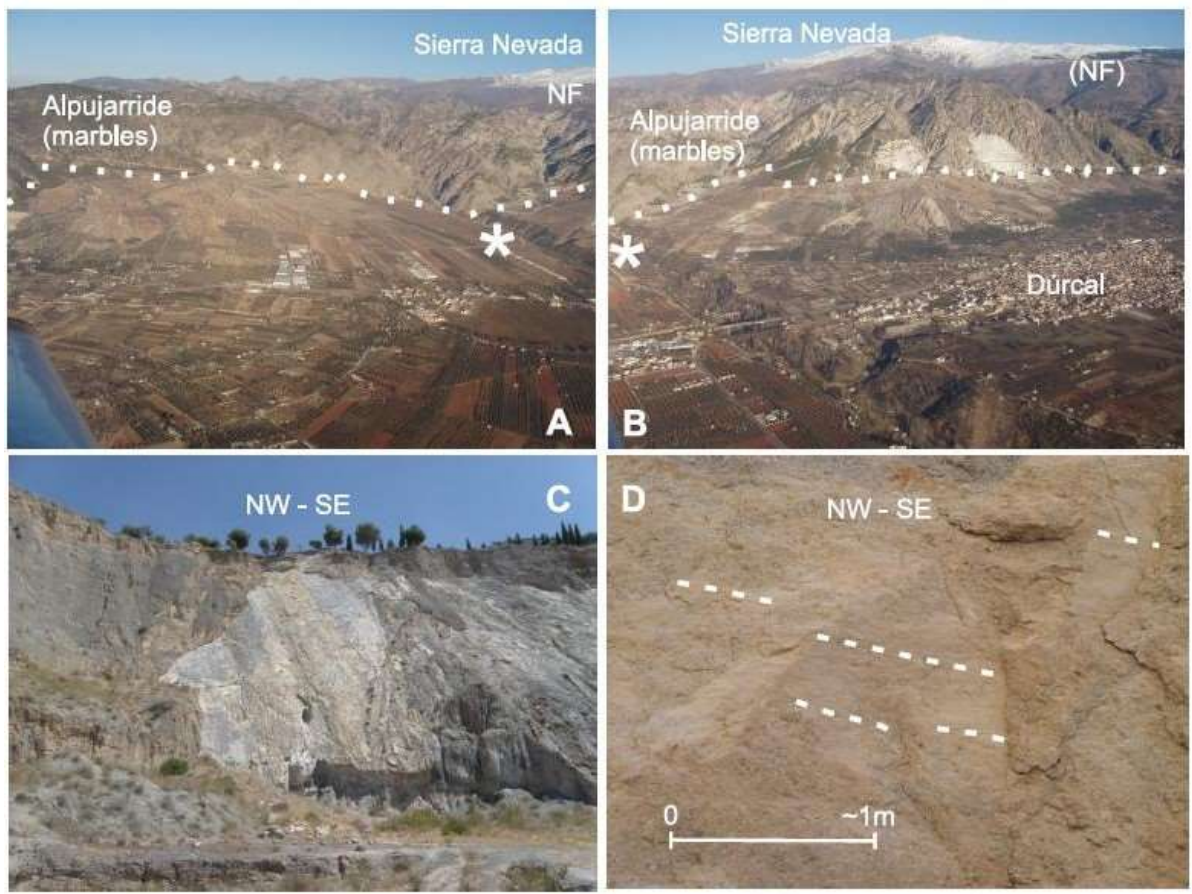

Fig. 4. Field view of the Padul Fault in the segment of GPS network. Location in Fig. 2B.

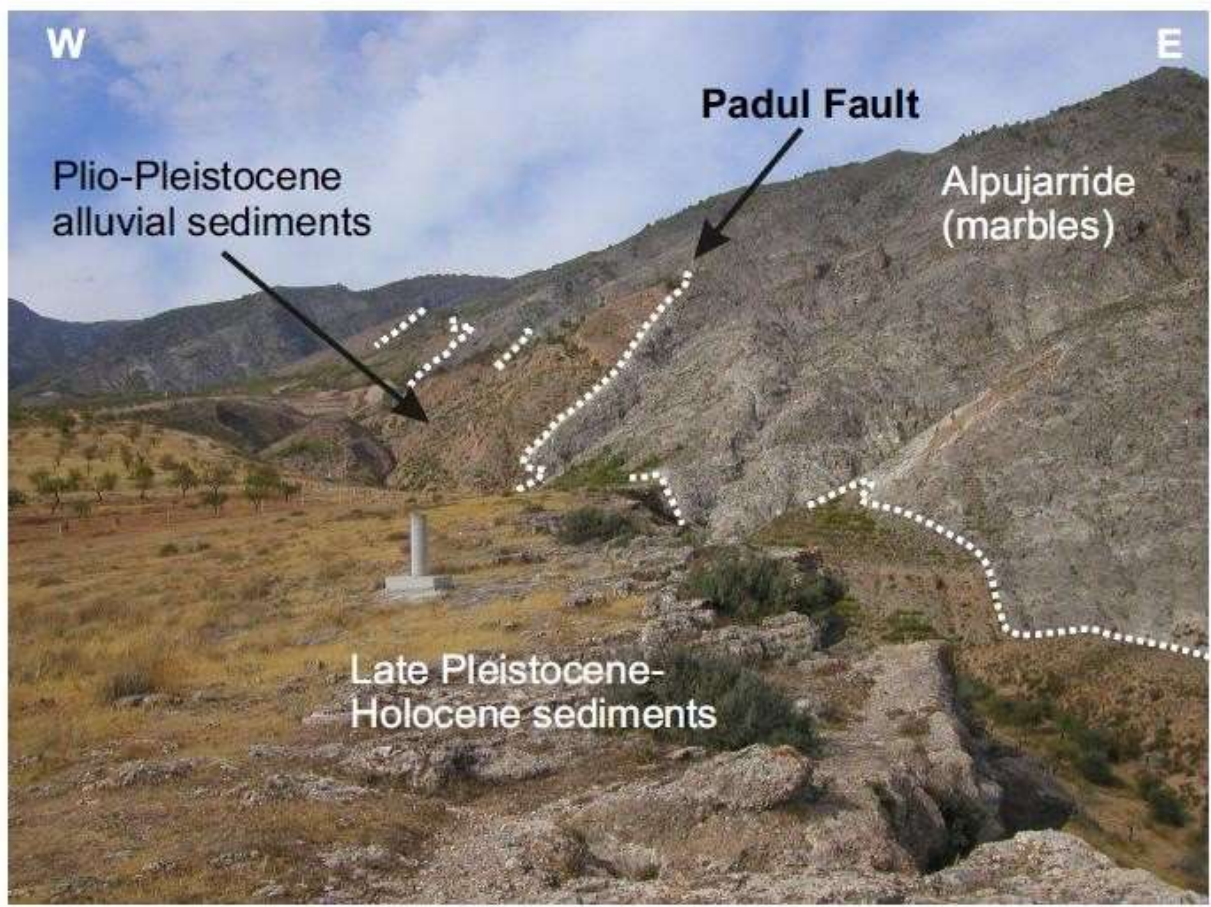

Fig. 5. E-W recent and active faults SW of Sierra Nevada. Faults affecting Plio-Quaternary conglomerates at the base of the sedimentary infill (A) and small half grabens in Plio-Quaternary sediments (B). Locations in Figs. 1B and 2A respectively.

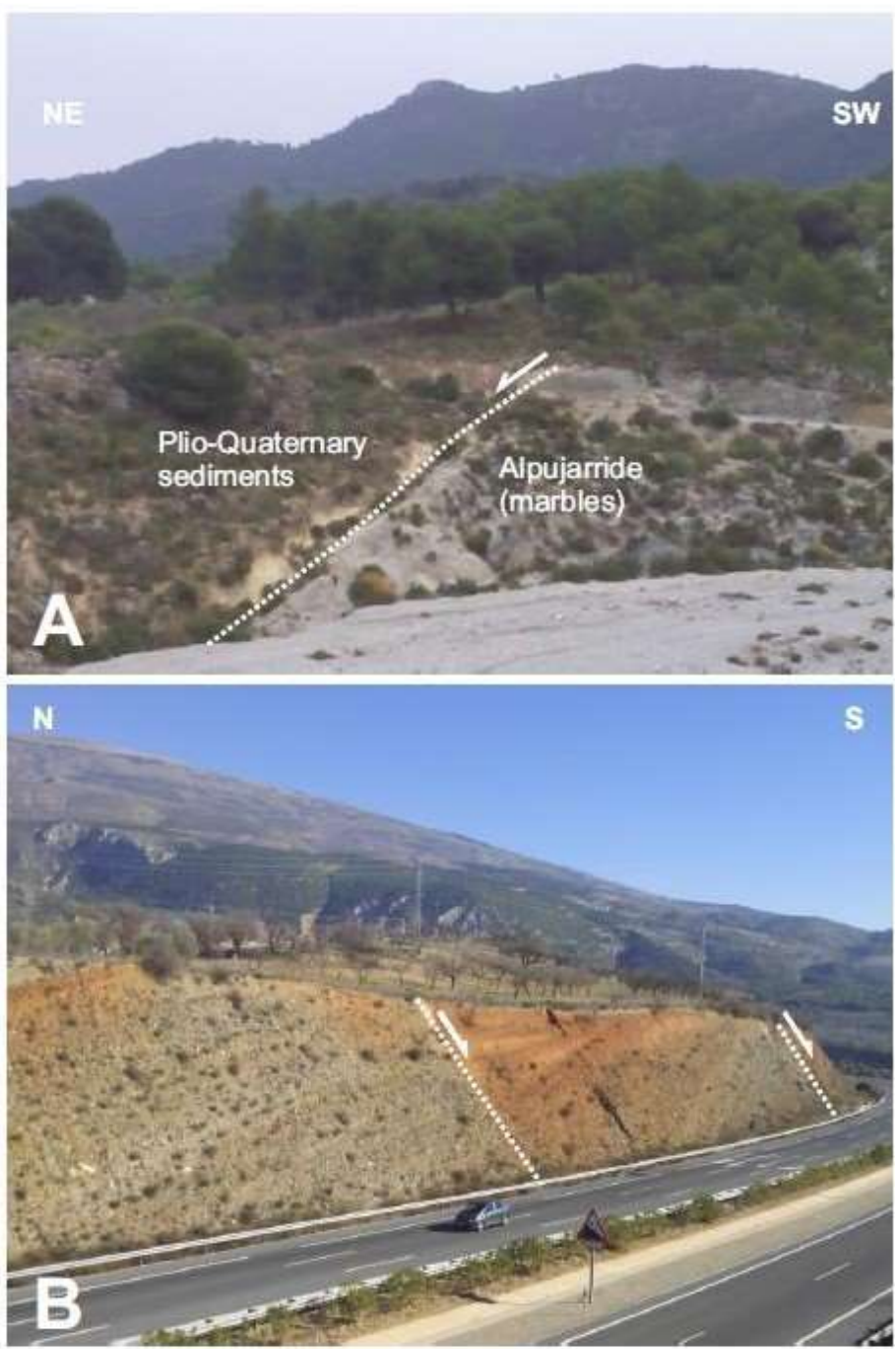

Fig. 6. GPS Position time series (in meters). The 3PAD time series has an additional observation campaign in July 2001 from the Granada Basin monitoring network observed in that year (Gil et al., 2002). Slopes of the solid lines are obtained using the position in East and North components derived from the GIPSY processing.

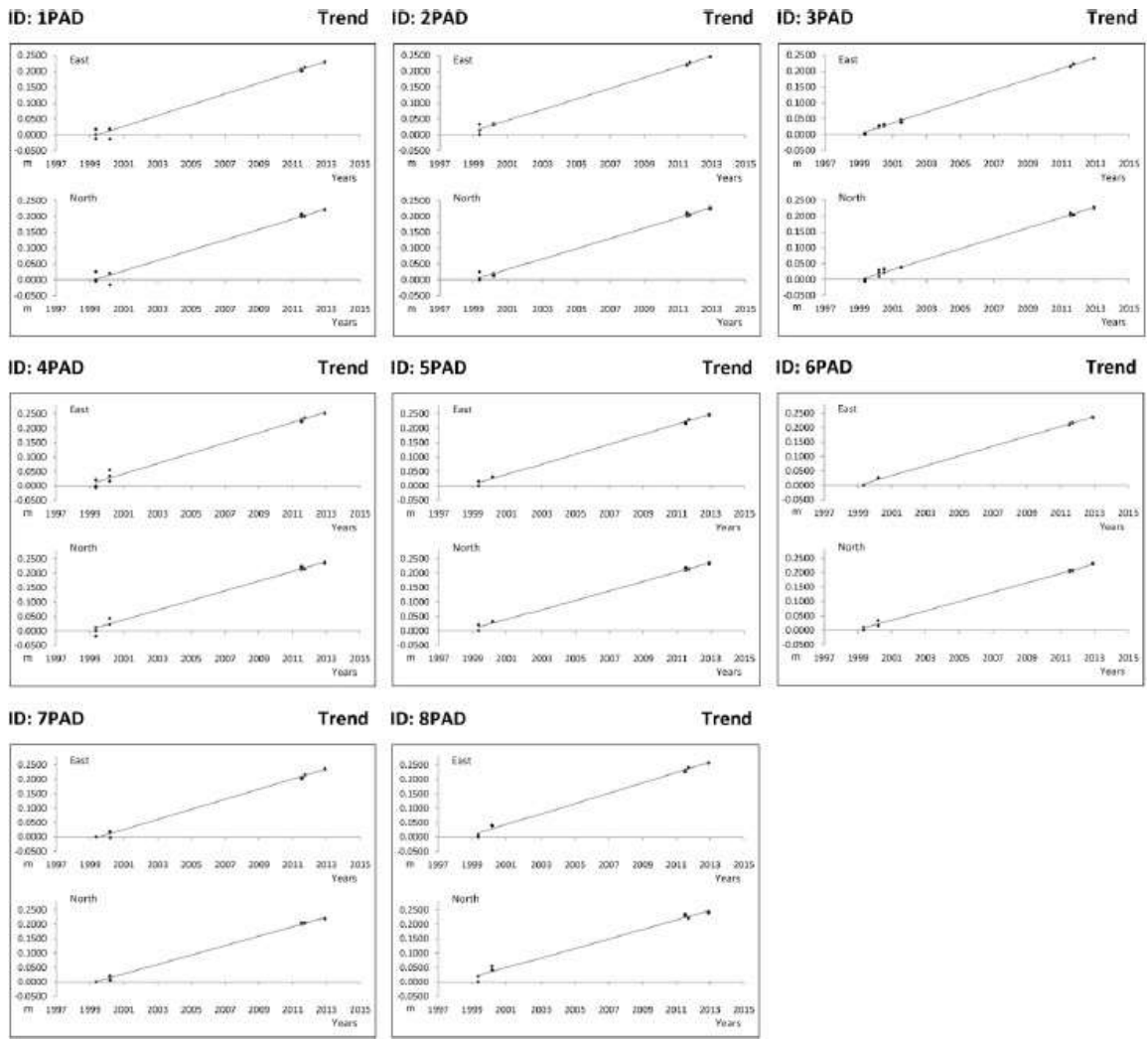

Fig. 7. GPS-derived velocity field in IGS08 reference frame and standard error ellipses.

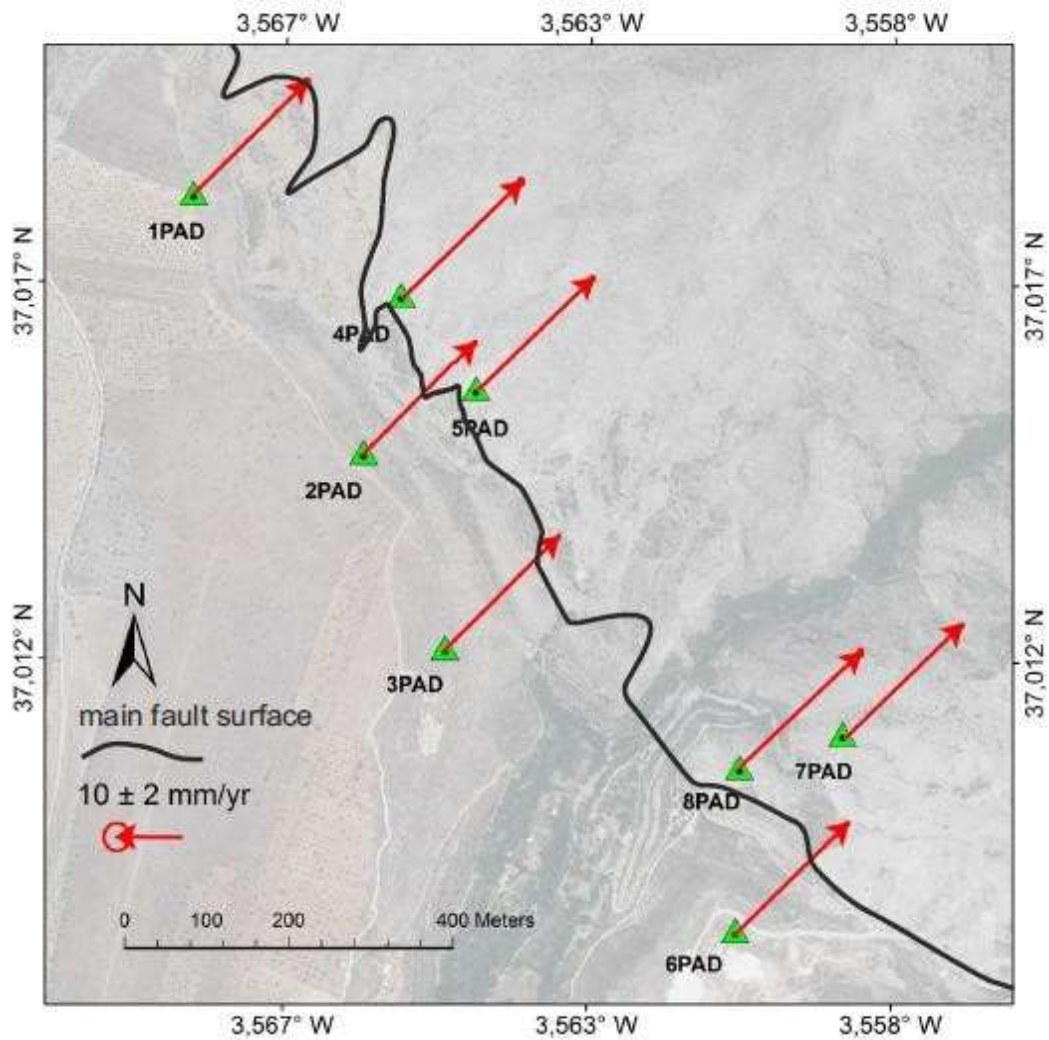

Fig. 8. GPS-derived horizontal velocity field with respect to average velocity of 4PAD, 5PAD, 7PAD and 8PAD sites (17.7 and 16.5 mm for East and North components respectively).

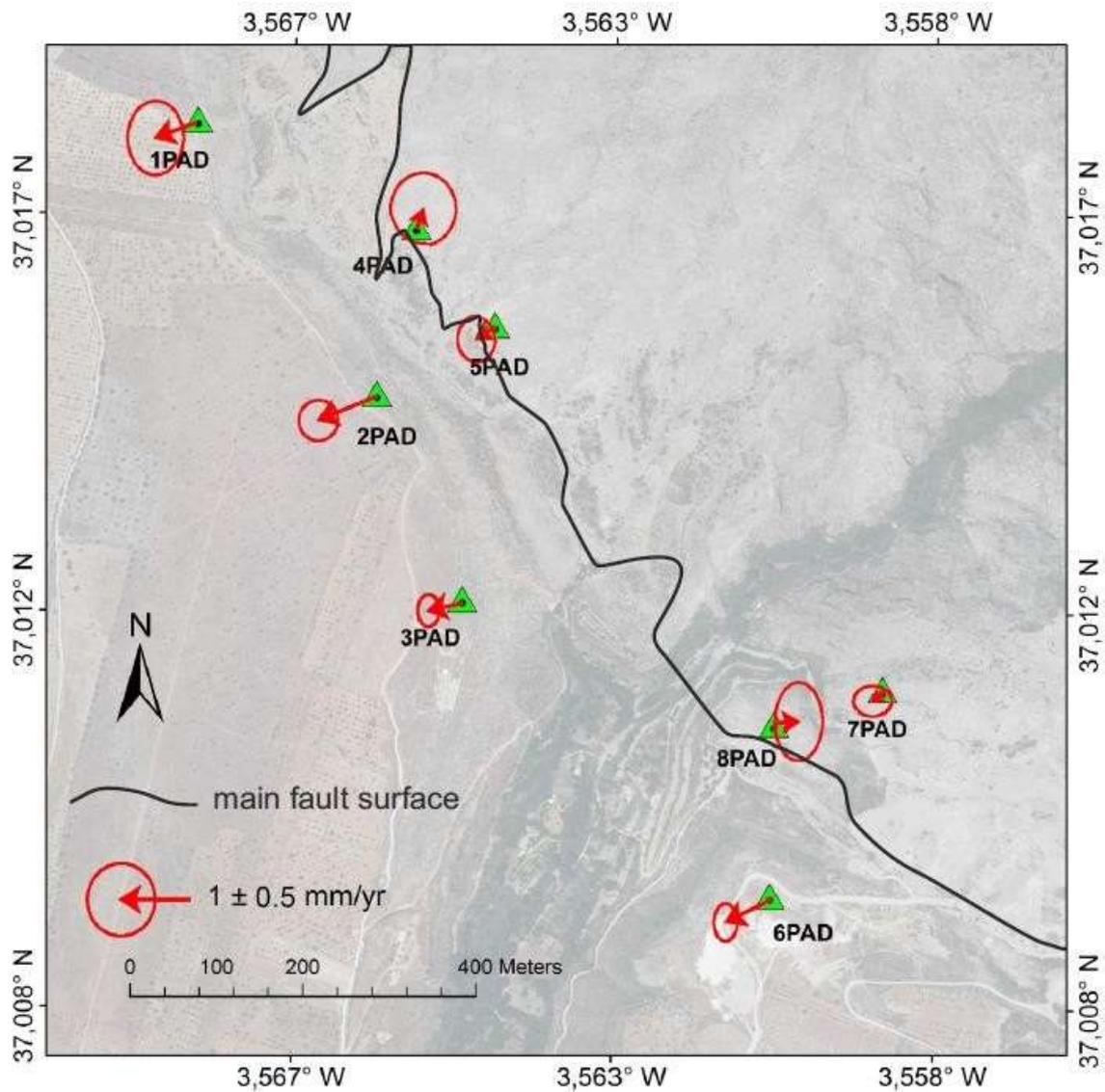

Fig. 9. The Padul Fault kinematics influenced by the Sierra Nevada Antiform. A, Active uplift of Sierra Nevada in the frame of the regional NW-SE compression and Padul Fault accommodating the NE-SW orthogonal extension. Northwestwards displacement of the Sierra produced a rotation of the regional stress field and the local stress field in the Padul Fault determined a normal sinistral kinematics, in agreement with most of the observed striae. B, The recent decreasing of northwestwards transport of the Sierra, evidenced by GPS data, determined a present-day normal to normal dextral kinematics according to the regional stress field.

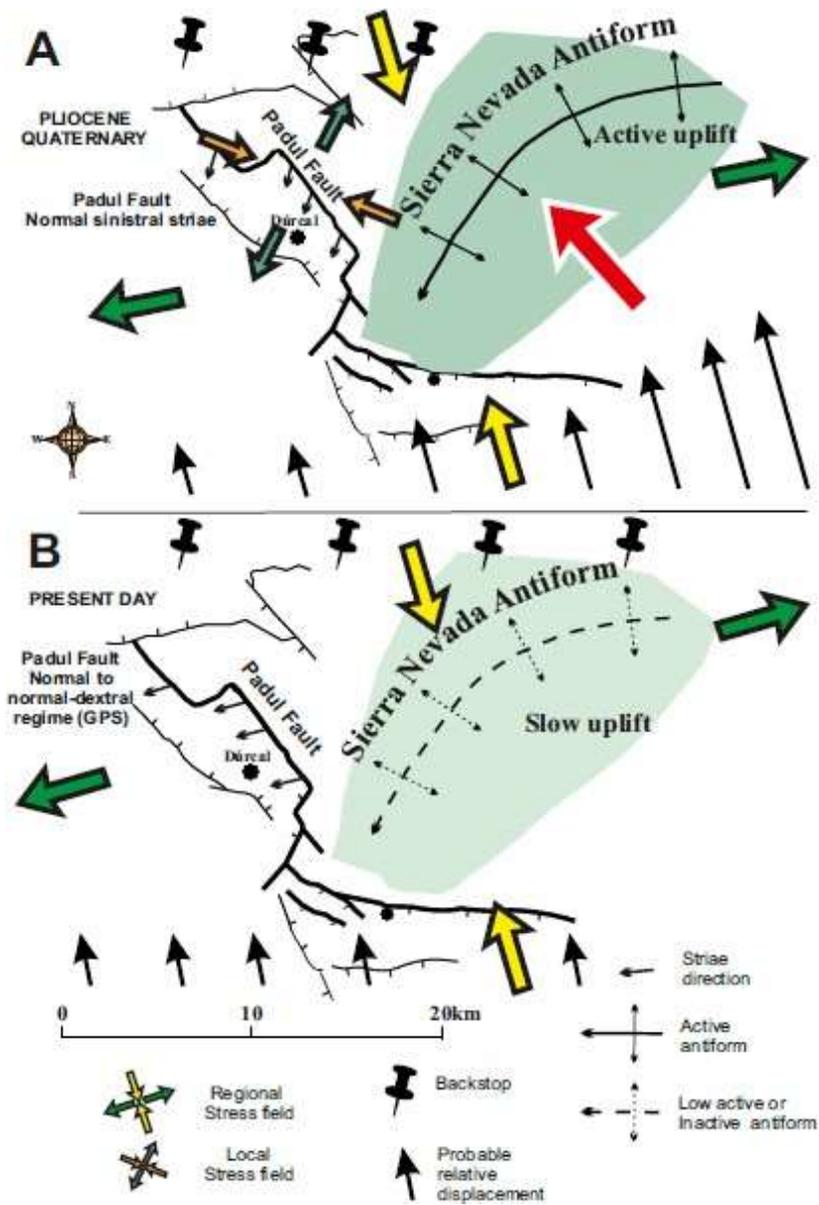

**Table captions**

Table 1. Site coordinates and GPS-derived horizontal velocities at the Padul Fault monitoring network in IGS08 reference frame. VE, VN: Velocities in East and North components.

| Coordinates | | | Absolute velocities (mm/yr) | | | |
|---|---|---|---|---|---|---|
| Site | Lat. (°N) | Long. (°E) | VE | $\sigma$E | VN | $\sigma$N |
| 1PAD | 39.017657 | -3.567929 | 17.1 | $\pm 0.4$ | 16.3 | $\pm 0.5$ |
| 2PAD | 37.014799 | -3.565582 | 16.9 | $\pm 0.3$ | 16.2 | $\pm 0.3$ |
| 3PAD | 37.012640 | -3.564447 | 17.2 | $\pm 0.2$ | 16.4 | $\pm 0.2$ |
| 4PAD | 37.016539 | -3.565072 | 17.8 | $\pm 0.5$ | 16.8 | $\pm 0.5$ |
| 5PAD | 37.015511 | -3.564049 | 17.4 | $\pm 0.3$ | 16.4 | $\pm 0.3$ |
| 6PAD | 37.009529 | -3.560453 | 17.1 | $\pm 0.2$ | 16.2 | $\pm 0.3$ |
| 7PAD | 37.011713 | -3.558993 | 17.5 | $\pm 0.3$ | 16.4 | $\pm 0.2$ |
| 8PAD | 37.011390 | -3.560405 | 18.0 | $\pm 0.3$ | 16.5 | $\pm 0.5$ |

Padul Fault is the main structure accomodating NE-SW extension in central Betics

GPS data determines a horizontal N66ºE extensional rate of 0.5 mm/yr

Earthquakes may reach up to M=6 but fault slips partially occurred by creep

Sierra Nevada Antiform interacts with Padul Fault determining its kinematics